\documentstyle[12pt]{article}
\def\scr{\cal}          
\def\ptl{\partial}
\def\out{\mathop{\rm out}}
\def\vac{\mathop{\rm vac}}
\def\ora#1{{\stackrel{\to}{#1}}}
\def\ola#1{{\stackrel{\gets}{#1}}}
\def\square{\mbox{{$\sqcup$}\llap{$\sqcap$}}}
\newcommand\bmphi{{\mbox{\boldmath$\phi$}}}
\newcounter{multilinear}
\renewcommand\theequation{\thesection.\arabic{equation}}
\begin{document}
\title{\bf The Quantized $O(1,2)\big/O(2)\times Z_2$ Sigma Model
Has No Continuum Limit\\
in Four Dimensions.\\
I\@. Theoretical Framework.
\\[-5cm]
{\footnotesize  hep-lat/9205014 \hfill 20 May 1992 \hfill UTREL 92-01}
\\[5cm]
}
\author{\sc Jorge de Lyra,\thanks{Permanent address: \
Universidade de S\~ao Paulo, Instituto de Fisica, Departamento de Fisica
Matem\'atica, C.P. 20516, 01498 S\~ao Paulo S.P., Brazil.}\quad
Bryce DeWitt, See Kit Foong,\thanks{Permanent address: \
Department of Physics, Faculty of Science, Ibaraki University,
Mito 310, Japan.}\\
\sc Timothy Gallivan, Rob Harrington, Arie Kapulkin,\\
\sc Eric Myers, {\rm and} Joseph Polchinski\\[\smallskipamount]
Center for Relativity and Theory Group\\
Department of Physics\\
The University of Texas at Austin\\
Austin, Texas \ 78712}
\date{}
\maketitle\baselineskip18pt
\begin{abstract}\baselineskip12pt
The nonlinear sigma model for which the field takes its values in the
coset space $O(1,2)\big/O(2)\times Z_2$ is similar to quantum gravity in
being perturbatively nonrenormalizable and having a noncompact curved
configuration space.
It is therefore a good model for testing nonperturbative methods that
may be useful in quantum gravity, especially methods based on lattice
field theory.
In this paper we develop the theoretical framework necessary for
recognizing and studying a consistent nonperturbative quantum field
theory of the $O(1,2)\big/O(2)\times Z_2$ model.
We describe the action, the geometry of the configuration space, the
conserved Noether currents, and the current algebra, and we construct a
version of the Ward-Slavnov identity that makes it easy to switch from a
given field to a nonlinearly related one.
Renormalization of the model is defined via the effective action and via
current algebra.
The two definitions are shown to be equivalent.
In a companion paper we develop a lattice formulation of the theory that
is particularly well suited to the sigma model, and we report the
results of Monte Carlo simulations of this lattice model.
These simulations indicate that as the lattice cutoff is removed the
theory becomes that of a pair of massless free fields.
Because the geometry and symmetries of these fields differ from those of
the original model we conclude that a continuum limit of the
$O(1,2)\big/O(2)\times Z_2$ model which preserves these properties does
not exist.
\end{abstract}

\section{Introduction}
Nonlinear sigma models in \mbox{4-dimensional} spacetime are not
perturbatively renormalizable.
They are therefore often viewed as being at best phenomenological,
incapable of representing physical reality at a fundamental level.
However, perturbative nonrenormalizability has never
been proved to imply nonrenormalizability, i.e.,
the impossibility of giving precise meaning to a model, even in the
extreme ultraviolet limit, in some nonperturbative sense.
The purpose of this and the paper following is to present evidence that
it is indeed impossible to give such meaning to at least one model: \
the $O(1,2)\big/O(2)\times Z_2$ sigma model.

Interest in this model stems from the following features that it shares
with quantum gravity:
\begin{enumerate}
\item Each theory has a single scale constant $\mu$ having the
dimensions of mass.
\item Each has a noncompact curved configuration space.
\item Corresponding Feynman graphs for each (in perturbation theory)
have identical degrees of divergence.
\item Neither is perturbatively renormalizable.
\end{enumerate}
Despite the negative result, experience gained in studying the
$O(1,2)\big/O(2)\times Z_2$ model is useful in approaching the more
difficult problem of quantum gravity, not only in regard to methodology,
but also in setting ranges for certain parameters and in discovering the
limits of present supercomputers.
Although a positive result, i.e., demonstration of the existence of a
continuum limit for the $O(1,2)\big/O(2)\times Z_2$ model, would have
been of obvious interest, no conclusion can yet be drawn about the
existence or nonexistence of a continuum limit for quantum gravity.
In no part of our analysis have we found reasons to have expected a
negative result {\sl a~priori}.
Each theory requires separate investigation, and it must be stressed
that quantum gravity differs in several important respects from a sigma
model:
\begin{enumerate}
\item Its invariance group, the diffeomorphism group, is local, whereas
the invariance group $O(1,2)$ of the sigma model is global.
\item The ground state of the sigma model is degenerate, and spontaneous
symmetry breaking occurs.
There is no evidence that the vacuum state of quantum gravity is
degenerate.
\item Quantization of the gravitational field smears the light cone.
The light cone for the sigma model remains fixed and sharp.
\end{enumerate}

No method is presently known for studying nonlinear sigma models
nonperturbatively in the extreme ultraviolet other than that of direct
assault via the functional integral of Feynman.
In field theory no nonperturbative definition of the functional integral
is known other than that of taking the continuum limit of the
corresponding integral on a lattice.
In this paper we describe the continuum theory, pretending that it
exists, in order to motivate our later choice of lattice action and of
things to compute.
The theoretical framework that will guide the computations is given in
some detail because there are important differences between noncompact
sigma models and the more familiar compact ones.
For example, noncompact models undergo no transition to a phase of
unbroken symmetry.
In the following paper we introduce a lattice action that is
particularly well suited to models with curved configuration spaces, and
we describe the lattice Monte Carlo simulation of the
$O(1,2)\big/O(2)\times Z_2$ model, using this action.
\setcounter{equation}{0}
\section{Action}
When we pass to the lattice simulation on the computer we shall need to
work in Euclidean space, but for most of this paper we shall work in a
Minkowski space with metric $\left(\eta_{\mu\nu}\right)=
\left(\eta^{\mu\nu}\right)=\mathop{\rm diag}({-1},1,1,1)$.
The action functional of a general sigma model has the form
\begin{equation}
S=-{\textstyle{1\over 2}}\mu^2\int g_{ij}\left(\phi(x)\right)
\phi^i{}_{,\mu}\,\phi^j{}_,{}^\mu\,d^4x\,,\label{2.1}
\end{equation}
where commas followed by Greek indices denote differentiation with
respect to the Minkowski coordinates $x^\mu({\mu=0},1,2,3)$ and can be
raised and lowered by means of the Minkowski metric.
The $\phi^i$ are scalar fields that take their values in charts of
configuration space, and the $g_{ij}$ are the components
(in those charts) of the metric tensor of configuration space.
The configuration space itself is always a coset space of a Lie group,
and the metric tensor is the natural group invariant one.
The configuration space is therefore a symmetric space with a
covariantly constant curvature tensor $R_{ijkl}$ and a strictly constant
curvature scalar $R\left(=R_{ij}{}^{ij}\right)$.
Only those coset spaces are chosen for which the natural metric is
positive definite.

The fields $\phi^i$ will be taken dimensionless.
In units with $\hbar=c=1$ the constant $\mu$ has then the dimensions of
mass.
It is sometimes convenient to absorb the constant $\mu$ into the fields
by redefining
\begin{equation}
\psi^i=\mu\phi^i\,,\label{2.2}
\end{equation}
and to follow the more standard convention of regarding scalar fields in
four dimensions as having the dimensions of mass.
This redefinition is useful, for example, in a chart in configuration
space in which the metric takes the form
\begin{equation}
g_{ij}=\delta_{ij}-\textstyle{1\over 3}R^0_{ikjl}\phi^k
\phi^l+\cdots\,,\label{2.3}
\end{equation}
$R^0_{ikjl}$ being the curvature tensor at $\phi=0$ in this chart.
The action (\ref{2.1}) becomes
\begin{equation}
S=-{\textstyle{1\over 2}}\int\left(\delta_{ij}
   -\textstyle{1\over 3}\mu^{-2}R^0_{ikjl}\psi^k\psi^l+\cdots\,\right)
\psi^i{}_{,\mu}\,\psi^j{}_,{}^\mu\,d^4x\,,\label{2.4}
\end{equation}
and the role of {\sl coupling constant\/} in the theory is seen to be
played by the inverse quantity~$\mu^{-2}$.
Of course, the Lagrangian of the actions (\ref{2.1}) and (\ref{2.4}) is
generally nonpolynomial, and there is in fact an infinity of coupling
constants: \ the powers of~$\mu^{-2}$.
We prefer to view $\mu$ as simply a scale parameter for the theory.
In recognition of the model's superficial resemblance to quantum gravity
we shall call $\mu$ the {\sl bare Planck mass}.

Many years ago Palais and Mostow\footnote{
R.~S. Palais, {\sl J.~Math.\ Mech.} {\bf6}, 673 (1957).
G.~D. Mostow, {\sl Annals of Math.} {\bf65}, 432 (1957).}
showed that coset spaces of semisimple Lie groups can always be embedded
in vector spaces in such a way that the group actions on each coset
space can be represented as linear homogeneous transformations of the
corresponding embedding space.
This means that the field variables $\phi^i$ can be replaced by a larger
set of variables $\phi^a$ together with a set of Lagrange multipliers
$\lambda^A$ that enforce the embedding constraints.
The $\lambda^A$ themselves remain invariant under the actions of the
group.
Since only a single chart is needed in the embedding space it is often
convenient to present nonlinear sigma models in this form.

The $O(1,2)\big/O(2)\times Z_2$ model is defined, in this form, by the
action
\begin{eqnarray}
\bar S&=&-{\textstyle{1\over 2}}\mu^2\int
\left[\eta_{ab}\,\phi^a{}_{,\mu}\,\phi^b{}_,{}^\mu+\lambda
\left(\eta_{ab}\,\phi^a\phi^b+1\right)\right]\,d^4x,
\qquad a,b\in\{0,1,2\}\,,\label{2.5}\\[\medskipamount]
\left(\eta_{ab}\right)&=&\mathop{\rm diag}
(-1,1,1)\,,\label{2.6}\\[\medskipamount]
\noalign{\noindent\rm together with the constraint\medskip}
\phi^0&\ge&1\,.\label{2.7}
\end{eqnarray}
The embedding space is seen to be a \mbox{3-dimensional} Minkowski space
on which the actions of $O(1,2)$ are Lorentz transformations.
The Lagrange multiplier enforces the constraint
\begin{equation}
\eta_{ab}\,\phi^a\phi^b=-1\label{2.8}
\end{equation}
which, together with (\ref{2.7}), identifies the configuration space as
the upper sheet of a spacelike hyperboloid in the embedding space.

The group invariant metric on the hyperboloid is that induced by the
metric $\eta_{ab}$ of the embedding space.
If one parameterizes the configuration space by
\begin{eqnarray}
\phi^0&=&\cosh s\,,\nonumber\\[\medskipamount]
\phi^1&=&\sinh s\cos\theta\,,\nonumber\\[\medskipamount]
\phi^2&=&\sinh s\sin\theta\,,\label{2.9}
\end{eqnarray}
then one may write the induced line element in the form
\begin{equation}
-\left(d\phi^0\right)^2+\left(d\phi^1\right)^2
+\left(d\phi^2\right)^2=ds^2+\sinh^2s\,d\theta^2\,,\label{2.10}
\end{equation}
and the action (\ref{2.1}) becomes
\begin{equation}
S=-{\textstyle{1\over 2}}\mu^2\int
\left(s_{,\mu}\,s_,{}^\mu+\sinh^2s\,\theta_{,\mu}\,\theta_,{}^\mu\right)
\,d^4x\,.\label{2.11}
\end{equation}
The coordinates $s$, $\theta$ are here allowed to cover the
configuration space an infinite number of times so that the
{\sl fields\/} $s$, $\theta$ may be taken differentiable wherever the
$\phi^a$ are differentiable.
\setcounter{equation}{0}
\section{Geometry of the configuration space}
The configuration space of the $O(1,2)\big/O(2)\times Z_2$ model has
constant negative curvature and is noncompact.
In fact it is topologically $\rm I\!R^2$ and can itself be covered by a
single chart, for example by the components of the \mbox{2-vector}
\begin{eqnarray}
\bmphi&=&\left(\matrix{\phi^1\cr
\noalign{\medskip}
\phi^2}\right)\,.\label{3.1}\\[\medskipamount]
\noalign{\noindent\rm The constraints (\ref{2.7}) and (\ref{2.8}) can be
expressed by\medskip}
\phi^0&=&\sqrt{1+\bmphi^2}\,,\label{3.2}
\end{eqnarray}
and it is not difficult to verify that the metric in this chart takes
the form
\begin{equation}
g_{ij}=\delta_{ij}-{\phi^i\phi^j\over 1+\bmphi^2},
\qquad i,j\in\{1,2\}\,.\label{3.3}
\end{equation}
The curvature tensor is also easily derived:
\begin{eqnarray}
R_{ijkl}&=&-\left(g_{ik}\,g_{jl}
                 -g_{il}\,g_{jk}\right)\,,\label{3.4}\\[\medskipamount]
R_{ij}&=&R_{ikj}{}^k=-g_{ij}\,,\label{3.5}\\[\medskipamount]
R&=&R_i{}^i=-2\,.\label{3.6}
\end{eqnarray}

Most nonlinear sigma models that have been studied have compact
configuration spaces.
We shall see later that noncompactness of its configuration space leads
to profound differences in the behavior of the
$O(1,2)\big/O(2)\times Z_2$ model from that of the compact models.
This will reflect itself in the need to deal with the lattice simulation
of the model by methods that will in part be unfamiliar.
We note at this stage just one fact: \
The area of a large circle in the $O(1,2)\big/O(2)\times Z_2$
configuration space increases not quadratically but
{\sl exponentially\/} with radius.
This means that the functional integral must probe an increasingly huge
region of the space of field histories as the continuum limit is
approached.

We shall need later an expression for the geodetic distance $\Delta$, in
the configuration space, between two points having coordinates
$(s,\theta)$ and $(s',\theta')$ respectively.
The simplest way to compute this is to note first that $s$ is the
geodetic distance from the point $\bmphi=0$ to the point $(s,\theta)$.
Therefore, since the geodetic distance between two points remains
invariant under the actions of $O(1,2)$, one may apply a Lorentz boost
that brings one of the two points to $\bmphi=0$.
This is most easily done in terms of the variables $\phi^a$,
$a\in\{0,1,2\}$, and one finds
\begin{eqnarray}
\Delta&=&\cosh^{-1}
\left(-\eta_{ab}\,\phi^a\phi'{}^b\right)\,.\label{3.7}\\[\medskipamount]
\noalign{\noindent\rm Substitution of expressions (\ref{2.9}) into
(\ref{3.7}) yields\medskip}
\Delta&=&\cosh^{-1}\left[\cosh s\cosh s'
-\sinh s\sinh s'\cos(\theta-\theta')\right]\,.\label{3.8}
\end{eqnarray}
\setcounter{equation}{0}
\section{Conserved currents}
The infinitesimal actions of the invariance group on the variables
$\phi^i$ may be expressed in the form
\begin{equation}
\delta\phi^i=Q^i{}_\alpha(\phi)\delta\xi^\alpha\,,\label{4.1}
\end{equation}
where the $\delta\xi^\alpha$ are infinitesimal group parameters and the
$Q^i{}_\alpha$ are the components (in the configuration space chart) of
a set of vector fields ${\bf Q}_\alpha$ on the configuration space,
which satisfy the Lie bracket relations
\begin{equation}
\left[{\bf Q}_\alpha,{\bf Q}_\beta\right]
    =-{\bf Q}_\gamma\,c^\gamma{}_{\alpha\beta}\,,\label{4.2}
\end{equation}
the $c^\gamma{}_{\alpha\beta}$ being the structure constants of the
group.
The statement that the Lagrangian
\begin{equation}
L=-\textstyle{1\over 2}\mu^2g_{ij}\,\phi^i{}_{,\mu}
                                  \,\phi^j{}_,{}^\mu\label{4.3}
\end{equation}
is invariant under (\ref{4.1}) is just the statement that $g_{ij}$ is
group invariant:
\begin{equation}
{\scr L}_{{\bf Q}_\alpha}\,g_{ij}=0\,.\label{4.4}
\end{equation}

Using (\ref{4.1}) together with the field equations
\begin{equation}
{\ptl L\over\ptl\phi^i}-\left(
{\ptl L\over\ptl\phi^i{}_{,\mu}}\right)_{,\mu}=0\,,\label{4.5}
\end{equation}
one may rewrite the invariance condition in the form
\begin{equation}
  0=\delta L={\ptl L\over\ptl\phi^i}
\delta\phi^i+{\ptl L\over\ptl\phi^i{}_{,\mu}}
\delta\phi^i{}_{,\mu}=j_\alpha{}^\mu{}_{,\mu}\,
\delta\xi^\alpha\label{4.6}
\end{equation}
where the $j_\alpha{}^\mu$ are the Noether currents:
\begin{equation}
j_\alpha{}^\mu={\ptl L\over\ptl\phi^i_{,\mu}}Q^i{}_\alpha
=-\mu^2g_{ij}\,Q^i{}_\alpha\,\phi^j{}_,{}^\mu\,.\label{4.7}
\end{equation}

Expression (\ref{4.7}) is valid for any sigma model.
It is completely equivalent to the corresponding expression obtained
when the alternative action is used, involving the linear embedding
variables $\phi^a$ and the Lagrange multipliers~$\lambda_A$.
In this case the infinitesimal group transformation law takes the form
\begin{eqnarray}
\delta\phi^a&=&G^a{}_{\alpha b}\,\phi^b
\delta\xi^\alpha\label{4.8}\\[\medskipamount]
\noalign{\noindent\rm or, with Latin indices suppressed,\medskip}
\delta\phi&=&G_\alpha\,\phi\delta\xi^\alpha\,,\label{4.9}
\end{eqnarray}
where the matrices $G_\alpha$ satisfy the commutation relation
\begin{equation}
\left[G_\alpha,G_\beta\right]=
G_\gamma\,c^\gamma{}_{\alpha\beta}\,.\label{4.10}
\end{equation}
Since the Lagrange multipliers are invariant under the group they play
no role in the definition of the currents, and one finds
\begin{equation}
j_\alpha{}^\mu={\ptl\bar L\over\ptl\phi_{,\mu}}
G_\alpha\,\phi\label{4.11}
\end{equation}
where $\bar L$ is the Lagrangian for the alternative action.
Expression (\ref{4.11}) is usually more convenient to
use than (\ref{4.7}).

Since the constants $\delta\xi^\alpha$ in eq.~(\ref{4.6}) are arbitrary
we have
\begin{equation}
j_\alpha{}^\mu{}_{,\mu}=0\,,\label{4.12}
\end{equation}
which implies conservation of the total charges
\begin{equation}
q_\alpha=\int_\Sigma j_\alpha{}^\mu\,d\Sigma_\mu\,.\label{4.13}
\end{equation}
Here $\Sigma$ is any smooth deformation of a global spacelike Cauchy
hypersurface in spacetime and $d\Sigma_\mu$ is its surface element.

In the case of the $O(1,2)\big/O(2)\times Z_2$ model it is convenient to
replace the single group index $\alpha$ by a pair of indices $ab$, so
that equation~(\ref{4.8}) takes the simple form
\begin{equation}
\delta\phi^a=\textstyle{1\over 2}G^a{}_{cd\,b}\,\phi^b\delta\xi^{cd}=
\delta\xi^a{}_b\,\phi^b\,.\label{4.14}
\end{equation}
Here we have
\begin{equation}
G^a{}_{cd\,b}=\delta^a{}_c\,\eta_{db}-\delta^a{}_d\,\eta_{cb},
\qquad\delta\xi^a{}_b=\eta_{bc}\,\delta\xi^{ac}\,.\label{4.15}
\end{equation}
The group parameters $\delta\xi^{ab}$ are those of an infinitesimal
Lorentz transformation in the configuration space and satisfy
\begin{equation}
\delta\xi^{ab}=-\delta\xi^{ba}\,.\label{4.16}
\end{equation}
Reference to eqs.~(\ref{2.5}) and (\ref{4.11}) shows that the currents
are
\begin{eqnarray}
j_{ab}{}^\mu&=&-\mu^2\eta_{cd}\,\phi^{c\mu}_,G^d{}_{ab\,e}\,\phi^e
=\mu^2\left(\phi_a\,\phi_{b,}{}^\mu-\phi_b\,\phi_{a,}{}^\mu\right)
\,,\label{4.17}\\[\medskipamount]
\noalign{\noindent\rm where $\phi_a=\eta_{ab}\phi^b$.
If the hypersurface $\Sigma$ is chosen to be $x^0=\rm constant$ then the
corresponding charges can be expressed in the form\medskip}
q_{ab}&=&\int j_{ab}{}^0\,d^3{\bf x}=-\mu^2\int
\left(\phi_a\,\phi_{b,0}-\phi_b\,\phi_{a,0}\right)
\,d^3{\bf x}\,.\label{4.18}
\end{eqnarray}
\setcounter{equation}{0}
\section{Functional integrals. Ward-Slavnov identities.}
Since we shall ultimately attempt to define the quantum theory of the
$O(1,2)\big/O(2)\times Z_2$ model by means of the functional integral of
Feynman, it is appropriate that we discuss quantum problems in
functional-integral language from the outset.
In principle the functional integral gives a direct representation of
transition amplitudes:
\setcounter{multilinear}{1}
\renewcommand\theequation{\thesection.\arabic{equation}\alph{multilinear}}
\begin{eqnarray}
\left<\out\mid\mathop{\rm in}\right>&=&
N\int e^{iS[\phi]}\mu[\phi]\,d\phi\label{5.1a}\\[\medskipamount]
\addtocounter{equation}{-1} \addtocounter{multilinear}{1}
&=&\bar N\int e^{i\bar S[\phi,\lambda]}\,d\phi\,d\lambda\,.\label{5.1b}
\end{eqnarray}
Here $\left|\mathop{\rm in}\right>$ and $\left|\out\right>$ are vectors
corresponding to arbitrary ``in'' and ``out'' states, the integrations
are over field histories satisfying boundary conditions appropriate to
those states, and it is understood that expressions (\ref{2.1}) and
(\ref{2.5}) for $S$ and $\bar S$ may need to be made more precise by
imposition of integration limits as well as addition of boundary terms.
$N$~and $\bar N$ are normalization constants and $\mu[\phi]$ is the
configuration space measure:
\renewcommand\theequation{\thesection.\arabic{equation}}
\begin{equation}
\mu[\phi]=\prod_xg^{1/2}\left(\phi(x)\right),\qquad
g=\det\left(g_{ij}\right)\,.\label{5.2}
\end{equation}
In principle a measure functional $\bar\mu[\phi,\lambda]$ should be
inserted in the integrand of (\ref{5.1b}), but it would be simply a
constant because the embedding space is flat.
The volume elements of (\ref{5.1a}) and (\ref{5.1b}) are respectively
\setcounter{multilinear}{1}
\renewcommand\theequation{\thesection.\arabic{equation}\alph{multilinear}}
\begin{eqnarray}
d\phi&=&\prod_{i,x}d\phi^i(x)\,,\label{5.3a}\\[\medskipamount]
\addtocounter{equation}{-1} \addtocounter{multilinear}{1}
d\phi\,d\lambda&=&\prod_x\,\prod_ad\phi^a(x)
\prod_Ad\lambda_A(x)\,.\label{5.3b}
\end{eqnarray}
Although expressions (\ref{5.2}), (\ref{5.3a}) and (\ref{5.3b}) are
purely formal they have a ready interpretation on a lattice.

The functional integral may also be used to represent ``in-out'' matrix
elements of chronologically ordered operators:
\setcounter{multilinear}{1}
\begin{eqnarray}
\left<\out\right|T\left(A[\phi]\right)\left|\mathop{\rm in}\right>&=&
N\int A[\phi]e^{iS[\phi]}\mu[\phi]\,d\phi\label{5.4a}\\[\medskipamount]
\addtocounter{equation}{-1} \addtocounter{multilinear}{1}
&=&\bar N\int A[\phi]e^{i\bar S[\phi,\lambda]}d\phi
\,d\lambda\,.\label{5.4b}
\end{eqnarray}
The symbol $\phi$ on the left of these equations is to be understood as
a quantum operator, on the right as an integration variable.

Suppose now the dummy variables of integration $\phi^i$ in
eq.~(\ref{5.4a}) are replaced by
\renewcommand\theequation{\thesection.\arabic{equation}}
\begin{equation}
\phi'{}^i=\phi^i+Q^i{}_\alpha(\phi)\delta\xi^\alpha\,,\label{5.5}
\end{equation}
where the $\delta\xi^\alpha$, instead of being constants as in
eq.~(\ref{4.1}), are now scalar functions of compact support in
spacetime.
The replacement (\ref{5.5}) leaves (\ref{5.4a}) unchanged, but it
produces the following changes in the individual factors under the
integral sign:
\begin{eqnarray}
\delta A[\phi]&=&\int{
\delta A[\phi]\over\delta\phi^i(x)}Q^i{}_\alpha\left(\phi(x)\right)
\delta\xi^\alpha(x)\,d^4x\,,\label{5.6}\\[\bigskipamount]
\delta e^{iS[\phi]}&=&ie^{iS[\phi]}\int\left[
{\ptl L\over\ptl\phi^i}Q^i{}_\alpha\,\delta\xi^\alpha+
{\ptl L\over\ptl\phi^i_{,\mu}}
\left(Q^i{}_\alpha\,\delta\xi^\alpha\right)_{,\mu}\right]
\,d^4x\nonumber\\[\medskipamount]
&=&ie^{iS[\phi]}\int j_\alpha{}^\mu\delta\xi^\alpha_{,\mu}
\,d^4x\nonumber\\[\medskipamount]
&=&-ie^{iS[\phi]}\int j_\alpha{}^\mu{}_{,\mu}\,\delta\xi^\alpha
\,d^4x\,,\label{5.7}\\[\bigskipamount]
\delta\left(\mu[\phi]\,d\phi\right)&=&\mu[\phi']
   \,d\phi'-\mu[\phi]\,d\phi\nonumber\\[\medskipamount]
&=&\left\{\mu[\phi']-\mu[\phi]+\mu[\phi']
   \left[\ptl(\phi')\big/\ptl(\phi)-1\right]\right\}
\,d\phi\nonumber\\[\medskipamount]
&=&\mu[\phi]\,d\phi\sum_xg^{-1/2}\left[
\left(\ptl g^{1/2}\!\big/\ptl\phi^i\right)Q^i{}_\alpha+g^{1/2}
   \ptl Q^i{}_\alpha\big/\ptl\phi^i\right]\delta\xi^\alpha\,.\label{5.8}
\end{eqnarray}
The second line of (\ref{5.7}) follows from the observation that the
Lagrangian would be invariant under (\ref{5.5}) if the
$\delta\xi^\alpha$ were constants.
Expression (\ref{5.8}) actually vanishes by virtue of the following
immediate corollary of eq.~(\ref{4.4}):
\begin{equation}
0={\scr L}_{{\bf Q}_\alpha}\,g^{1/2}=\ptl
\left(g^{1/2}Q^i{}_\alpha\right)\big/\ptl\phi^i\,.\label{5.9}
\end{equation}

Taking note of the arbitrariness of the $\delta\xi^\alpha$ one may
therefore infer
\begin{eqnarray}
0&=&N\int\left\{{\delta A[\phi]\over\delta\phi^i(x)}Q^i{}_\alpha
  \left(\phi(x)\right)-iA[\phi]j^\mu{}_{,\mu}(x)\right\}e^{iS[\phi]}
\mu[\phi]\,d\phi\nonumber\\[\medskipamount]
&=&\left<\out\right|T\left({\delta A[\phi]\over\delta\phi^i(x)}
Q^i{}_\alpha\left(\phi(x)\right)-iA[\phi]j^\mu{}_{,\mu}(x)\right)
\left|\mathop{\rm in}\right>\,.\label{5.10}
\end{eqnarray}
Because the ``in'' and ``out'' states are themselves arbitrary one can,
in fact, infer the operator identity
\begin{equation}
{\ptl\over\ptl x^\mu}T\left(A[\phi]j^\mu(x)\right)=-iT
\left({\delta A[\phi]\over\delta\phi^i(x)}Q^i{}_\alpha
\left(\phi(x)\right)\right)\,.\label{5.11}
\end{equation}
Note here that, because the functional integral of the difference of two
functionals is the difference of their individual integrals, the
chronological ordering that eq.~(\ref{5.4a}) defines commutes with the
operation of differentiation with respect to the spacetime coordinates.
Equation~(\ref{5.11}) is a slightly generalized version of the standard
{\sl Ward-Slavnov identity}.

The Ward-Slavnov identity can be extended to the embedding variables
$\phi^a$ by working with expression (\ref{5.4b}) instead of (\ref{5.4a})
and replacing (\ref{5.5}) by
\begin{equation}
\phi'=\left(1+G_\alpha\,\delta\xi^\alpha\right)\phi\,.\label{5.12}
\end{equation}
The invariance of the volume element $d\phi$ under this transformation
follows from
\begin{equation}
\mathop{\rm tr}G_\alpha=0\,,\label{5.13}
\end{equation}
which in turn follows from the semisimplicity of the invariance group.
The extended Ward-Slavnov identity takes the form
\begin{equation}
{\ptl\over\ptl x^\mu}T\left(A[\phi]j_\alpha{}^\mu(x)\right)=-iT\left(
 {\delta A[\phi]\over\delta\phi(x)}G_\alpha\,\phi(x)\right)
\,.\label{5.14}
\end{equation}
\setcounter{equation}{0}
\section{Current algebra}
Setting $A[\phi]=1$ in eq.~(\ref{5.14}) one gets the operator
conservation law
\begin{equation}
j_\alpha{}^\mu{}_{,\mu}=0\,.\label{6.1}
\end{equation}
On the other hand, choosing $A[\phi]=\phi(x')$, writing
\begin{equation}
T\left(\phi(x')j_\alpha{}^\mu(x)\right)=\theta\left(x'{}^0-x^0
\right)\phi(x')j_\alpha{}^\mu(x)+\theta\left(x^0-x'{}^0\right)
               j_\alpha{}^\mu(x)\phi(x')\label{6.2}
\end{equation}
where $\theta$ is the step function, and making use of (\ref{6.1}), one
gets
\begin{eqnarray}
\left[\phi(x'),j_\alpha{}^0(x)\right]\delta\left(x'{}^0-x^0\right)
&=&iG_\alpha\,\phi(x)\delta(x'-x)\,,\label{6.3}\\[\medskipamount]
\noalign{\noindent\rm which is equivalent to the equal-time
commutator\medskip}
\left[\phi(x),j_\alpha{}^0(x')\right]&=&iG_\alpha\,\phi(x)\delta
({\bf x}-{\bf x}'),\qquad x^0=x'{}^0\,.\label{6.4}\\[\medskipamount]
\noalign{\noindent\rm Integration of this equation over ${\bf x}'$
yields\medskip}
\left[\phi,q_\alpha\right]&=&iG_\alpha\,\phi\label{6.5}
\end{eqnarray}
which, by virtue of the group invariance of the Lagrangian $\bar L$, in
turn yields
\begin{eqnarray}
\left[{\ptl\bar L\over\ptl\phi_{,\mu}},q_\alpha\right]
 &=&-i{\ptl\bar L\over\ptl\phi_{,\mu}}G_\alpha
\,,\label{6.6}\\[\medskipamount]
\noalign{\noindent\rm and hence\medskip}
\left[j_\alpha{}^\mu,q_\beta\right]&=&i{\ptl\bar L\over\ptl\phi_{,\mu}}
\left[G_\alpha,G_\beta\right]\phi=
ij_\gamma{}^\mu c^\gamma{}_{\alpha\beta}\,,\label{6.7}\\[\medskipamount]
\left[q_\alpha,q_\beta\right]
&=&iq_\gamma\,c^\gamma{}_{\alpha\beta}\,.\label{6.8}
\end{eqnarray}
Using (\ref{4.15}) one easily finds, for the $O(1,2)\big/O(2)\times Z_2$
model,
\begin{eqnarray}
\left[\phi^0(x),j_{ij}{}^0(x')\right]\delta
\left(x^0-x'{}^0\right)&=&0\,,\label{6.9}\\[\bigskipamount]
\left[\phi^k(x),j_{ij}{}^0(x')\right]\delta\left(x^0-x'{}^0\right)&=&i
\left[\delta^k{}_i\,\phi_j(x)-\delta^k{}_j\,\phi_i(x)\right]
\delta(x-x')\,,\label{6.10}\\[\bigskipamount]
\left[\phi^0(x),j_{0i}{}^0(x')\right]\delta\left(x^0-x'{}^0\right)
  &=&i\phi_i(x)\delta(x-x')\,,\label{6.11}\\[\bigskipamount]
\left[\phi^k(x),j_{0i}{}^0(x')\right]\delta\left(x^0-x'{}^0\right)
&=&-i\delta^k{}_i\,\phi_0(x)\delta(x-x')\nonumber\\[\medskipamount]
&=&i\delta^k{}_i\,\phi^0(x)\delta(x-x')\,,\label{6.12}
\end{eqnarray}
where $i,j,k\in\{1,2\}$.

It is sometimes convenient to use chart coordinates in configuration
space other than the~$\phi^i$.
For the $O(1,2)\big/O(2)\times Z_2$ model these will be chosen so as to
transform like the $\phi^i$ under the $O(2)$ subgroup.
The most general such coordinates are
\begin{equation}
\varphi^i=f\left(\,|\bmphi|\,\right)\phi^i\,,\label{6.13}
\end{equation}
where $f\left(\,|\bmphi|\,\right)$ is a positive-valued smooth function
such that $\left|\bmphi\right|f\left(\,|\bmphi|\,\right)$ has everywhere
positive slope.
For example, if
\begin{equation}
f\left(\,|\bmphi|\,\right)=
{\sinh^{-1}\left|\bmphi\right|\over|\bmphi|}\label{6.14}
\end{equation}
then the $\varphi^i$ are the Riemann normal coordinates~$\sigma^i$:
\begin{equation}
\left.\begin{array}{l}\sigma^1=s\cos\theta\\[\medskipamount]
                      \sigma^2=s\sin\theta.\end{array}\right\}\label{6.15}
\end{equation}
It is readily verified that in general we have
\begin{eqnarray}
\left[\varphi^k(x),j_{ij}{}^0(x')\right]\delta\left(x^0-x'{}^0\right)&=&i
\left[\delta^k{}_i\,\varphi_j(x)-\delta^k{}_j\,\varphi_i(x)\right]
\delta(x-x')\,,\label{6.16}\\[\medskipamount]
\left[\varphi^k(x),j_{0i}{}^0(x')\right]\delta\left(x^0-x'{}^0\right)&=&
i{\ptl\varphi^k(x)\over\ptl\phi^i(x)}\phi_0(x)
\delta(x-x')\,.\label{6.17}
\end{eqnarray}
\setcounter{equation}{0}
\section{Vacuum and 1-particle states}
The classical energy corresponding to the action (\ref{2.1}) is
nonnegative and vanishes for constant fields.
Each different constant field corresponds to a different classical
vacuum state.
Since $O(1,2)$ acts transitively on the configuration space the
classical vacua are obtainable from one another by group operations.
One expects the degeneracy of the classical ground state to be reflected
in a corresponding degeneracy of the quantum ground state.
If $\left|\vac\right>$ is the vector corresponding to one of the quantum
vacua then the vectors corresponding to the others are obtained through
multiplication by unitary operators
$\exp\left(iq_\alpha\,\xi^\alpha\right)$.

Since all points of configuration space are equivalent under $O(1,2)$
one can always choose a Lorentz frame in the embedding space so that the
classical vacuum state is $\phi^1=\phi^2=0$.
The corresponding quantum vacuum is fixed by the conditions
\begin{equation}
\bar\phi^1=\bar\phi^2=0,\qquad
\bar\phi^a=\left<\vac\right|\phi^a\left|\vac\right>,
\qquad a\in\{0,1,2\}\,.\label{7.1}
\end{equation}
Here the normalization
\begin{equation}
\left<\vac\mid\vac\right>=1\label{7.2}
\end{equation}
will be assumed.
Each quantum vacuum will also be assumed to be stable.

The degeneracy of the quantum ground state implies that the basic
particles of the theory are two Goldstone bosons.
The masslessness of these particles is already suggested by the form of
the first term in the expansion (\ref{2.4}) of the classical action.
Since the classical action generates no bare vertices having fewer than
four prongs, perturbation theory, if it were valid, would imply that
these particles, despite being massless, are stable.\footnote{Phase
space restrictions forbid the decay of a single massless particle into
three or more others.
The derivative coupling also suppresses the decay.}
Since perturbation theory fails one must either {\sl assume\/} that
these particles are stable or else assume that \mbox{2-point} functions
exist and that the quantum vacuum, like the classical vacuum, has an
$O(2)$ invariance that is never broken.
In the latter case Goldstone's theorem will guarantee that these
particles are stable, for, as will be shown later, the degeneracy of the
quantum ground state is never removed.

Since the \mbox{$S$-matrix} of the theory is expected to be independent
of the choice of interpolating field one should in principle be able to
construct particle creation and annihilation operators out of any of the
fields (\ref{6.13}) including $\phi^i$ itself.
Asymptotic arguments imply that the creation operators are given by
\begin{equation}
a^i({\bf p})^*=i\bar\mu_\varphi\int_{\pm\infty}u(x,{\bf p})
\left({\ora\ptl\over\ptl x_\mu}-{\ola\ptl\over\ptl x_\mu}\right)
\varphi^i(x)\,d\Sigma_\mu\,,\label{7.3}
\end{equation}
where ``${+\infty}$'' and ``${-\infty}$'' denote Cauchy hypersurfaces in
the remote future and past respectively, either asymptotic region being
usable because of the stability of the particles.
Here $u$ is a {\sl mode function\/} that should strictly have the form
of a wave packet but which, if the packet is big enough, may effectively
be represented by a plane wave:
\begin{equation}
\left.\begin{array}{cc}\multicolumn{2}{c}{u(x,{\bf p})=
(2\pi)^{-3/2}(2\omega)^{-1/2}e^{ip\cdot x}},\\[\medskipamount]
\omega=p^0=\left|{\bf p}\right|,&p^2=0.\end{array}\right\}\label{7.4}
\end{equation}
In the Appendix it is shown that if the \mbox{1-particle} state vectors
are defined by
\begin{eqnarray}
\left|i,{\bf p}\right>
 &=&a^i({\bf p})^*\left|\vac\right>\label{7.5}\\[\medskipamount]
\noalign{\noindent\rm then the normalization\medskip}
\left<i,{\bf p}\mid j,{\bf p}'\right>&=&
\delta_{ij}\,\delta({\bf p}-{\bf p}')\label{7.6}
\end{eqnarray}
is secured by choosing the coefficient $\bar\mu_\varphi$ in (\ref{7.3})
to be the square root of the reciprocal of the residue at the ``pole''
at $p^2=0$ in the Fourier transform of the \mbox{2-point} function:
\begin{equation}
\left<\vac\right|T\left(\varphi^i(x)\varphi^j(x')\right)
\left|\vac\right>=-{i\over(2\pi)^4}\int
{\delta_{ij}\,e^{ip\cdot(x-x')}\over\bar\mu_\varphi{}^2
\left(p^2-i0\right)+\cdots}\,d^4p\,.\label{7.7}
\end{equation}
If $\varphi^i$ were replaced by $\phi^i$ and if all the terms in the
integrand of (\ref{2.4}) other than the first were missing, so that the
theory were ``free,'' then $\bar\mu_\varphi$ would be simply the bare
Planck mass~$\mu$.
For the nonlinear theory it must be computed.
This, in fact, is a major goal of the numerical simulation.

Strictly speaking, the singularity of the integrand of (\ref{7.7}) at
$p^2=0$ cannot be a pole when the massless particles interact but must
be a branch point.
The coefficient $\bar\mu_\varphi{}^2$ can nevertheless be identified
because the unwritten terms ``\ldots'' in the denominator of the
integrand generally vanish more rapidly than $p^2$ as $p^2\to 0$.
For example, in perturbation theory the first unwritten term behaves
like $p^6\ln p^2$.

The proof in the Appendix, of the normalization (\ref{7.6}), makes use
of the mode-function orthonormality relations
\begin{equation}
\left.\begin{array}{l}
\displaystyle-i\int_\Sigma u(x,{\bf p})^*\left({\ora\ptl\over\ptl x_\mu}
-{\ola\ptl\over\ptl x_\mu}\right)u(x,{\bf p}')\,d\Sigma_\mu
=\delta({\bf p}-{\bf p}')\\[\bigskipamount]
\displaystyle-i\int_\Sigma u(x,{\bf p})\left({\ora\ptl\over\ptl x_\mu}
-{\ola\ptl\over\ptl x_\mu}\right)u(x,{\bf p}')\,d\Sigma_\mu
=0,\end{array}\right\}\label{7.9}
\end{equation}
and of the defining condition
\begin{equation}
a^i({\bf p})\left|\vac\right>=0,\qquad
\left<\vac\right|a^i({\bf p})^*=0\,,\label{7.8}
\end{equation}
for the vacuum state vector.
Equations~(\ref{7.3}) and (\ref{7.9}), together with the completeness of
the mode functions, imply that $\varphi^i(x)$ may, in the remote past or
future, be effectively represented by
\begin{equation}
\varphi^i(x)=\int
\left[u(x,{\bf p})a^i({\bf p})+u(x,p)^*a^i({\bf p})^*\right]
\,d^3{\bf p},\qquad x^0\to\pm\infty\,.\label{7.10}
\end{equation}
This in turn implies
\begin{equation}
\left<\vac\right|\varphi^i(x)\left|\vac\right>=0\label{7.11}
\end{equation}
for $x^0\to\pm\infty$.
Because the action is an even functional of the fields eq.~(\ref{7.11})
in fact holds for all $x$ (cf.~eq.~(\ref{7.1})).
\setcounter{equation}{0}
\section{Renormalization via the effective action}
Renormalization is most easily discussed in terms of the effective
action, denoted here by $\Gamma\left[\bar\phi\right]$ or
$\bar\Gamma\left[\bar\phi,\bar\lambda\right]$ according as the classical
action is taken to be $S[\phi]$ or $\bar S[\phi,\lambda]$.
Consider first~$\bar\Gamma$.
The traditional way to define it is to introduce sources $J_a$, $J_A$
coupled to the $\phi^a$ and $\lambda^A$ respectively and to set
\begin{equation}
e^{i\bar W[J]}=\left<\out,\vac\mid\mathop{\rm in},\vac\right>=\bar N\int
e^{i\left[\bar S[\phi,\lambda]+\int\left(J_a\,\phi^a+J_A\,\lambda^A\right)
\,d^4x\right]}\,d\phi\,d\lambda\,,\label{8.1}
\end{equation}
where, because of the presence of the sources
(which are assumed to have compact support in spacetime), one must now
distinguish between ``in'' and ``out'' vacua.
The functional $\bar W[J]$ is used to define the following quantities:
\begin{equation}
\bar\phi^a={\delta\bar W\over\delta J_a},\qquad\bar\lambda^A
          ={\delta\bar W\over\delta J_A}\,,\label{8.2}
\end{equation}
\begin{equation}
G^{a_1\ldots a'_rA''_1\ldots A'''_s}={\delta\over\delta J_{a_1}(x)}
                               \cdots{\delta\over\delta J_{a_r}(x')}\,
{\delta\over\delta\lambda_{A_1}(x'')}\cdots
{\delta\over\delta\lambda_{A_s}(x''')}\bar W\,.\label{8.3}
\end{equation}
It is straightforward to show that these quantities are related to the
so called \mbox{$n$-point} functions:
\begin{eqnarray}
\left<\phi^a\right>&=&\bar\phi^a,\qquad
\left<\lambda^A\right>=\bar\lambda^A\,,\label{8.4}\\[\medskipamount]
\left<\phi^a\phi^{b'}\right>
&=&\bar\phi^a\bar\phi^{b'}-iG^{ab'},\mbox{ etc.}\,,\label{8.5}
\end{eqnarray}
the \mbox{$n$-point} functions themselves being special cases of the
general average
\begin{eqnarray}
\left<A[\phi,\lambda]\right>&=&{\left<\out,\vac\right|T
\left(A[\phi,\lambda]\right)\left|\mathop{\rm in},\vac\right>
          \over\left<\out,\vac\mid\mathop{\rm in},\vac\right>}
\nonumber\\[\medskipamount]
&=&{\int A[\phi,\lambda]e^{i\left[\bar S[\phi,\lambda]+\int
\left(J_a\,\phi^a+J_A\,\lambda^A\right)\,d^4x\right]}\,d\phi\,d\lambda
\over\int e^{i\left[\bar S[\phi,\lambda]+\int
\left(J_a\,\phi^a+J_A\,\lambda^A\right)\,d^4x\right]}\,d\phi\,d\lambda}
\,.\label{8.6}
\end{eqnarray}
$\bar\phi^a$~and $\bar\lambda^A$ are called {\sl mean fields\/} and the
$G$'s are known as {\sl correlation functions}.

The effective action is derived from the functional $\bar W[J]$ by the
Legendre transformation
\begin{equation}
\bar\Gamma\left[\bar\phi,\bar\lambda\right]=\bar W[J]
-\int\left(J_a\,\bar\phi^a+J_A\,\bar\lambda^A\right)\,d^4x\,.\label{8.7}
\end{equation}
It is not difficult to verify that $\bar\Gamma$ satisfies the equations
\begin{equation}
{\delta\bar\Gamma\over\delta\bar\phi^a}=-J_a,\qquad
{\delta\bar\Gamma\over\delta\bar\lambda^A}=-J_A\,.\label{8.8}
\end{equation}
Functional differentiation of these equations yields
\begin{equation}
\int\left({\delta^2\bar\Gamma\over\delta\bar\phi^a\delta\bar\phi^{c''}}
G^{c''b'}+{\delta^2\bar\Gamma\over\delta\bar\phi^a\delta\bar\lambda^{A''}}
G^{A''b'}\right)\,d^4x''=-\delta_a{}^b\delta(x-x'),
\mbox{ etc.}\,,\label{8.9}
\end{equation}
which reveals $G^{ab'}$, $G^{aB'}$, $G^{Ab'}$, $G^{AB'}$ as Green's
functions of the second functional derivative of~$\bar\Gamma$.
These are the {\sl full propagators\/} of the theory.
Further functional differentiation of eqs.~(\ref{8.9}) enables one to
express the higher order correlation functions in terms of these
propagators together with the functional derivatives of $\bar\Gamma$ of
order three and higher, which are known as the
{\sl full vertex functions}.
The resulting relations have a tree graph structure that allows one to
recognize $\bar\Gamma$ as the generator of the \mbox{1-particle}
irreducible amplitudes of the theory.

The effective action $\Gamma\left[\bar\phi\right]$ associated with the
classical action $S[\phi]$ is constructed in a different way.
The reason for this is that although the variables $\phi^a$ transform
linearly under the invariance group the variables $\phi^i$ transform
nonlinearly among themselves.
The direct coupling of the latter variables to sources would lead to a
$\Gamma$ which has no simple transformation law under the group.
What one does instead is first note that the above construction of
$\bar\Gamma$ is completely equivalent to defining it implicitly
(or recursively) by
\begin{equation}
e^{i\bar\Gamma\left[\bar\phi,\bar\lambda\right]}=\bar N\int\exp i
\left\{\bar S[\phi,\lambda]+\int\left[{\delta\bar\Gamma
\left[\bar\phi,\bar\lambda\right]\over\delta\bar\phi^a}
\left(\bar\phi^a-\phi^a\right)+{\delta\bar\Gamma
\left[\bar\phi,\bar\lambda\right]\over\delta\bar\lambda^A}
\left(\bar\lambda^A-\lambda^A\right)\right]\,d^4x\right\}
\,d\phi\,d\lambda\,.\label{8.10}
\end{equation}
{}From this it is easy to see that $\bar\Gamma$ must be linear in
the~$\bar\lambda^A$:
\begin{equation}
\bar\Gamma\left[\bar\phi,\bar\lambda\right]=
\hat\Gamma\left[\bar\phi\right]+\int\bar\lambda^A
       C_A\left[\bar\phi\right]\,d^4x\,.\label{8.11}
\end{equation}
One also notes that because the $\phi^a$, $\lambda^A$ transform linearly
under the group so do the $\bar\phi^a$,~$\bar\lambda^A$:
\begin{equation}
\delta\bar\phi^a=G^a{}_{\alpha b}\,\bar\phi^b\delta\xi^\alpha,\qquad
\delta\bar\lambda^A=0\,.\label{8.12}
\end{equation}
The sources $J_a$, $J_A$ may be assumed to transform contragradiently,
and $\hat\Gamma\left[\bar\phi\right]$ and the $C_A\left[\bar\phi\right]$
are therefore group invariant.

The effective action $\Gamma\left[\bar\phi\right]$ is now
constructed from $\bar\Gamma\left[\bar\phi,\bar\lambda\right]$ in the
same way as the classical action $S[\phi]$ is constructed from
$\bar S[\phi,\lambda]$.
One sets the sources $J_A$ equal to zero and solves the constraint
equations
\begin{equation}
{\delta\bar\Gamma\left[\bar\phi,\bar\lambda\right]\over\delta\bar\lambda^A}
       \equiv C_A\left[\bar\phi\right]=0\label{8.13}
\end{equation}
for the superfluous embedding variables (or rather their barred forms)
in terms of the chart variables in the coset space, and then substitutes
into $\hat\Gamma\left[\bar\phi\right]$
to get $\Gamma\left[\bar\phi\right]$.
In the case of the $O(1,2)\big/O(2)\times Z_2$ model, if the
$\bar\phi^a$ are constant fields (e.g., when the sources $J_a$ vanish)
equation~(\ref{8.13}) is necessarily equivalent to
\begin{equation}
\eta_{ab}\bar\phi^a\bar\phi^b=-Z^2\label{8.14}
\end{equation}
for some constant~$Z$.

If the model is to have a consistent continuum limit the effective
action must, because of symmetry requirements, reduce at low energies to
a local action of the classical form (\ref{2.1}), but with possibly a
new value of~$\mu$.
Thus
\begin{equation}
\Gamma\left[\bar\phi\right]=-{\textstyle{1\over 2}}\bar\mu^2\int\bar
g_{ij}\left(\bar\phi(x)\right)\bar\phi^i_{,\mu}\bar\phi^j_,{}^\mu
\,d^4x+\Delta\Gamma\left[\bar\phi\right]\,,\label{8.15}
\end{equation}
where $\Delta\Gamma$, which includes nonlocal contributions, becomes
important at higher energies.\footnote{The clean separation of $\Gamma$
into ``leading'' and ``subleading'' parts is implied by the same
power-counting arguments that show the theory to be perturbatively
nonrenormalizable.}
The metric $\bar g_{ij}$ is group invariant
(i.e., it satisfies eq.~(\ref{4.4})) but, because the $\bar\phi^a$ are
constrained by (\ref{8.14}) to a generally different hyperboloid than
the original classical variables $\phi^a$, $\bar g_{ij}$ is not equal to
the metric (\ref{3.3}) but is instead given by
\begin{equation}
\bar g_{ij}=
\delta_{ij}-{\bar\phi^i\bar\phi^j\over Z^2+\bar\bmphi^2}\,.\label{8.16}
\end{equation}
The original metric is restored if the quantum operator fields are
replaced by {\sl renormalized\/} fields:
\begin{equation}
\phi^a_R=Z^{-1}\phi^a\,.\label{8.17}
\end{equation}
The effective action then takes the form
\begin{equation}
\Gamma=-{\textstyle{1\over 2}}\mu_R{}^2\int g_{ij}
\left(\bar\phi_R(x)\right)\bar\phi^i_{R,\mu}\bar\phi^j_{R,}{}^\mu\,d^4x
+\Delta\Gamma\label{8.18}
\end{equation}
where
\begin{equation}
\mu_R=Z\bar\mu\,.\label{8.19}
\end{equation}
$\mu_R$~is the renormalized or ``experimentally observed'' Planck mass.
Note that because the Lagrangian is nonpolynomial the replacement
(\ref{8.17}) renormalizes an infinite number of vertex functions at
once.
Note also that although (\ref{8.17}) {\sl looks\/} like a standard
``wave function renormalization'' the renormalization constant $Z$ is
{\sl not\/} obtained from the residue of the pole of the \mbox{2-point}
function.
Rather, it is the constant $\bar\mu_\varphi$ of eq.~(\ref{7.7}) that is
defined by the residue.

When the sources $J_a$ vanish and the ``in'' and ``out'' vacua become
identical, the vacuum-fixing conditions (\ref{7.1}) imply
\begin{equation}
Z=\bar\phi^0=\left<\vac\right|\phi^0\left|\vac\right>\,,\label{8.20}
\end{equation}
\begin{equation}
\left<\vac\right|T\left(\phi^i_R(x)\phi^j_R(x')\right)
\left|\vac\right>=-iG^{ij'}_R\,,\label{8.21}
\end{equation}
(cf.~eq.~(\ref{8.5})).
$G^{ij'}_R$~is the Green's function of the second functional derivative
of the effective action with respect to the renormalized fields.
The structure of the local part of expression (\ref{8.18}) implies that
the low-energy behavior of this Green's function is given by
\begin{eqnarray}
G^{ij'}_R&=&{1\over(2\pi)^4}\int
{\delta_{ij}\,e^{ip\cdot(x-x')}\over\mu^2_R\left(p^2-i0\right)+\cdots}
\,d^4p\,.\label{8.22}\\[\medskipamount]
\noalign{\noindent\rm Comparison of eqs.~(\ref{7.7}), (\ref{8.21}) and
(\ref{8.22}) allows one to infer\medskip}
\mu_R&=&Z\bar\mu_\phi\,.\label{8.23}
\end{eqnarray}
Comparison of eq.~(\ref{8.19}) and (\ref{8.23}) in turn informs one that
the $\bar\mu$ of eq.~(\ref{8.15}) is just the $\bar\mu_\varphi$ of
eq.~(\ref{7.7}) when $\varphi=\phi$.
\setcounter{equation}{0}
\section{Renormalization via current algebra}
The renormalization analysis above makes special use of the variables
$\phi^a$, which transform linearly under the group.
Determination of the renormalized Planck mass $\mu_R$, however, should
not depend on which field variables are used.
The generality of eq.~(\ref{7.7}) suggests that the choice of variables
should, in fact, be irrelevant and that for every $\bar\mu_\varphi$
there should be an easily computable $Z_\varphi$ such that
\begin{equation}
\mu_R=Z_\varphi\,\bar\mu_\varphi\,.\label{9.1}
\end{equation}
In this section we show that this is indeed the case.

Consider the vacuum average $\left<\vac\right|T\left(\varphi^i
(x')j_{ab}{}^\mu(x)\right)\left|\vac\right>$ where $\varphi^i$ is
one of the fields (\ref{6.13}).
Because of the displacement invariance of the theory this average must
be a function of the differences $x'{}^\nu-x^\nu$.
It must also transform as a vector under Lorentz transformations.
Its Fourier transform must therefore have the form
\begin{equation}
\int e^{ip\cdot(x'-x)}
\left<\vac\right|T\left(\varphi^i(x')j_{ab}{}^\mu(x)\right)
\left|\vac\right>\,d^4x'=p^\mu X^i{}_{ab}(p^2)\label{9.2}
\end{equation}
where the $X^i{}_{ab}$ are certain scalar functions of~$p^2$.
These functions may be determined by multiplying eq.~(\ref{9.2}) by
$p_\mu$ and invoking the Ward-Slavnov identity (\ref{5.14}):
\begin{eqnarray}
p^2X^i{}_{ab}&=&-i\int
\left[{\ptl\over\ptl x'{}^\mu}e^{ip\cdot(x'-x)}\right]
\left<\vac\right|T\left(\varphi^i(x')j_{ab}{}^\mu(x)\right)
\left|\vac\right>\,d^4x'\nonumber\\[\medskipamount]
&=&-i\int e^{ip\cdot(x'-x)}{\ptl\over\ptl x^\mu}
\left<\vac\right|T\left(\varphi^i(x')j_{ab}{}^\mu(x)\right)
\left|\vac\right>\,d^4x'\nonumber\\[\medskipamount]
&=&-\int e^{ip\cdot(x'-x)}\delta(x'-x)
\left<\vac\right|{\ptl\varphi^i\over\ptl\phi^c}G^c{}_{abd}\,\phi^d
\left|\vac\right>\,d^4x'\nonumber\\[\medskipamount]
&=&-\left<\vac\right|\left({\ptl\varphi^i\over\ptl\phi^a}\phi_b
                          -{\ptl\varphi^i\over\ptl\phi^b}\phi_a
\right)\left|\vac\right>\,.\label{9.3}
\end{eqnarray}
The integration by parts and subsequent replacement of
$\ptl/\ptl x'{}^\mu$ by $-\ptl/\ptl x^\mu$, leading to the second line,
assumes that the vacuum average falls off sufficiently rapidly at
infinity.
Omission of the chronological ordering symbol in the last two lines is
allowed by the ultralocality of the operators involved.

$X^i{}_{ab}$~is seen to have a simple $1/p^2$ behavior.
The pole at $p^2=0$ can be removed by taking the Laplacian of the
original vacuum average and integrating by parts.
We do this for the special case $a=0$, $b=j$ and note that, because the
vacuum average depends only on the differences $x'{}^\nu-x^\nu$,
application of $\square'$ is equivalent to application of~$\square$:
\begin{eqnarray}
&&\int e^{ip\cdot(x'-x)}\square'
\left<\vac\right|T\left(\varphi^i(x')j_{0j}{}^\mu(x)\right)
\left|\vac\right>\,d^4x'\nonumber\\[\medskipamount]
&&\quad=-p^\mu p^2X^i{}_{0j}=p^\mu\left<\vac\right|
\left({\ptl\varphi^i\over\ptl\phi^0}\phi_j
     -{\ptl\varphi^i\over\ptl\phi^j}\phi_0\right)
\left|\vac\right>\nonumber\\[\medskipamount]
&&\quad=\delta_{ij}\,p^\mu\left<\vac\right|\left[f
\left(\,|\bmphi|\,\right)+{1\over 2}\left|\bmphi\right|f'
\left(\,|\bmphi|\,\right)\right]\phi^0\left|\vac\right>\,.\label{9.4}
\end{eqnarray}
The last line is obtained by invoking the explicit form (\ref{6.13}) and
the fact that the vacuum average must be proportional to $\delta_{ij}$
since, as far as the indices $i$ and $j$ are concerned it is an $O(2)$
invariant tensor.
This follows from the fact that the vacuum is $O(2)$ invariant, being a
zero-eigenvalue eigenstate of the charge~$q_{12}$.

Now consider the matrix element
$\left<\vac\right|j_{0j}{}^\mu(x)\left|i,{\bf p}\right>$.
Since $\left|\vac\right>$ and $\left|i,{\bf p}\right>$ are normalized
physical state vectors, and since $j_{ab}{}^\mu$ is the physical current
that yields the physically observable charges $q_{ab}$
(which generate group transformations), this matrix element must depend
only on observable renormalized quantities.
Moreover, its value must be independent of what field variables one uses
to calculate it.
Making use of eqs.~(\ref{7.3}), (\ref{7.5}), (\ref{7.8}) and
(\ref{9.4}), one gets
\begin{eqnarray}
&&\left<\vac\right|j_{0j}{}^\mu(x)\left|i,{\bf p}\right>
=i\bar\mu_\varphi\int_{-\infty}u(x',{\bf p})
\left({\ora\ptl\over\ptl x'_\nu}-{\ola\ptl\over\ptl x'_\nu}\right)
\left<\vac\right|j_{0j}{}^\mu(x)\varphi^i(x')
\left|\vac\right>\,d\Sigma'_\nu\nonumber\\[\medskipamount]
&&{}=-i\bar\mu_\varphi
\left(\int_{+\infty}-\int_{-\infty}\right)u(x',{\bf p})
\left({\ora\ptl\over\ptl x'_\nu}-{\ola\ptl\over\ptl x'_\nu}\right)
\left<\vac\right|T\left(\varphi^i(x')j_{0j}{}^\mu(x)\right)
\left|\vac\right>\,d\Sigma'_\nu\nonumber\\[\medskipamount]
&&{}=-i\bar\mu_\varphi(2\pi)^{-3/2}(2\omega)^{-1/2}e^{ip\cdot x}\int
e^{ip\cdot(x'-x)}\left(\ora\square{}'-\ola\square{}'\right)
\left<\vac\right|T\left(\varphi^i(x')j_{0j}{}^\mu(x)\right)
\left|\vac\right>\,d^4x'\nonumber\\[\medskipamount]
&&{}=-i\bar\mu_\varphi\left<\vac\right|\left[f
\left(\,|\bmphi|\,\right)+\textstyle{1\over 2}\left|\bmphi\right|f'
\left(\,|\bmphi|\,\right)\right]\phi^0
\left|\vac\right>\delta_{ij}\,p^\mu u(x,{\bf p})\,.\label{9.5}
\end{eqnarray}
In the special case $f\left(\,|\bmphi|\,\right)=1$, $\varphi^i=\phi^i$,
$\bar\mu_\varphi=\bar\mu_\phi=\bar\mu$, the final vacuum average is just
$Z$ (see eq.~(\ref{8.20})), and it follows from eq.~(\ref{8.23}) that
\begin{equation}
\left<\vac\right|j_{0j}{}^\mu(x)\left|i,{\bf p}\right>
=-i\mu_R\,\delta_{ij}\,p^\mu u(x,{\bf p})\label{9.6}
\end{equation}
showing that this matrix element does indeed depend only on observable
quantities.
Moreover, eq.~(\ref{9.1}) now follows, with the identification
\begin{equation}
Z_\varphi=\left<\vac\right|\left[f
\left(\,|\bmphi|\,\right)+\textstyle{1\over 2}\left|\bmphi\right|f'
\left(\,|\bmphi|\,\right)\right]\phi^0\left|\vac\right>\,.\label{9.7}
\end{equation}

This identification allows eq.~(\ref{8.21})
(with $G^{ij'}_R$ given by (\ref{8.22})) to be generalized:
\begin{eqnarray}
\left<\vac\right|T\left(\varphi^i_R(x)\varphi^j_R(x')\right)
\left|\vac\right>&=&-{i\over(2\pi)^4}\int
{\delta_{ij}\,e^{ip\cdot(x-x')}\over\mu^2_R\left(p^2-i0\right)+\cdots}
\,d^4p\,,\label{9.8}\\[\medskipamount]
\noalign{\noindent\rm where\medskip}
\varphi^i_R&=&Z^{-1}_\varphi\varphi^i\,.\label{9.9}
\end{eqnarray}
Note that since the fields $\phi^i_R$ and $\varphi^i_R$ are not
identical the unwritten terms in the denominators of the integrands of
expressions (\ref{8.22}) and (\ref{9.8}) will not generally be the same.
Only the coefficients of $p^2$ will be identical.

We now have two completely independent ways of computing~$\mu_R$.
One is to compute $Z_\varphi$ by (\ref{9.7}), carry out the field
renormalization (\ref{9.9}), and determine the residue of the pole at
$p^2=0$ in the Fourier transform of the \mbox{2-point} function
(\ref{9.8}).
The other is to compute the matrix element (\ref{9.6}).

The first method can immediately be translated to Euclidean space and
adapted for the computer.
The second requires a little modification.
Instead of working directly with the matrix element (\ref{9.6}) we
arrange for it to appear in a sum over intermediate states:
\begin{eqnarray}
&&\left<\vac\right|T\left(j_{0i}{}^\mu(x)j_{0j}{}^\nu(x')\right)
  \left|\vac\right>\nonumber\\[\medskipamount]
&&{}=\left<\vac\right|\left[\theta
\left(x^0-x'{}^0\right)j_{0i}{}^\mu(x)j_{0j}{}^\nu(x')+\theta
\left(x'{}^0-x^0\right)j_{0j}{}^\nu(x')j_{0i}{}^\mu(x)\right]
\left|\vac\right>\nonumber\\[\medskipamount]
&&{}=\sum_{k=1}^2\int\Big[\theta\left(x^0-x'{}^0\right)\left<\vac\right|
j_{0i}{}^\mu(x)\left|k,{\bf p}\right>\left<k,{\bf p}\right|
j_{0j}{}^\nu(x')\left|\vac\right>\nonumber\\[\medskipamount]
&&\phantom{=\sum_{k=1}^2\int\Big[}
+\theta\left(x'{}^0-x^0\right)\left<\vac\right|
j_{0j}{}^\nu(x')\left|k,{\bf p}\right>\left<k,{\bf p}\right|
j_{0i}{}^\mu(x)\left|\vac\right>\Big]\,d^3\bf p\nonumber\\
&&\phantom{=\sum_{k=1}^2\int\Big[}
+\cdots\,,\qquad\mu\not=\nu\,.\label{9.10}
\end{eqnarray}
Here we assume $\mu\not=\nu$ so that we may use the na\"\i ve definition of
chronological ordering without worrying about Schwinger terms.
In the final sum the omission of the vacuum as an intermediate state
follows from the Lorentz invariance and $O(2)$ invariance of the vacuum,
which implies
\begin{equation}
\left<\vac\right|j_{0i}{}^\mu(x)\left|\vac\right>=0\,.\label{9.11}
\end{equation}
The other unwritten terms in (\ref{9.10}) are sums over intermediate
states involving two or more particles.
These are expected to be of higher order in $p^2$ than the written term
when the Fourier transform is taken.

Fourier transforms will here be defined by
\begin{equation}
\tilde f(p)=(2\pi)^{-2}\int f(x)e^{ip\cdot x}\,d^4x\,.\label{9.12}
\end{equation}
To obtain the Fourier transform of (\ref{9.10}) first insert (\ref{9.6})
into the sum, obtaining
\begin{eqnarray}
&&\left<\vac\right|T\left(j_{0i}{}^\mu(x)j_{0j}{}^\nu(x')\right)
  \left|\vac\right>\nonumber\\[\medskipamount]
&&{}=(2\pi)^{-3}\mu^2_R\delta_{ij}\int{p^\mu p^\nu\over 2\omega}
\left[\theta\left(x^0-x'{}^0\right)e^{ip\cdot(x-x')}
     +\theta\left(x'{}^0-x^0\right)e^{ip\cdot(x'-x)}\right]
\,d^3{\bf p}+\cdots\nonumber\\[\medskipamount]
&&{}=-{i\over(2\pi)^4}\mu^2_R\delta_{ij}\int{p^\mu p^\nu\over p^2-i0}
e^{ip\cdot(x-x')}\,d^4p+\cdots\,,\qquad\mu\not=\nu\,,\label{9.13}
\end{eqnarray}
which immediately yields
\begin{equation}
\left<\vac\right|T\left(\tilde\jmath_{0i}{}^\mu(p)
\tilde\jmath_{0j}{}^\nu(p')\right)\left|\vac\right>=-i\mu^2_R
\delta_{ij}{p^\mu p^\nu\over p^2-i0}
\delta(p+p')+\cdots\,,\qquad\mu\not=\nu\,.\label{9.14}
\end{equation}
Since the unwritten terms are of higher order in $p^2$ it is evident
that $\mu^2_R$ may in principle be determined by examining (\ref{9.14})
in the limit $p^2\to 0$.
\setcounter{equation}{0}
\section{Euclidean space}
To pass to Euclidean space one makes the variable replacements
\begin{equation}
x^0=-ix^4,\qquad p_0=ip_4\,,\label{10.1}
\end{equation}
and carries out $90^\circ$ rotations in the complex planes of these
variables.
In the functional integrals (\ref{5.1a}) and (\ref{5.4a}), as well as in
expressions (\ref{5.2}) and (\ref{5.3a}) for the functional volume
element, the $x^\mu$ are simply labels, or generalized indices, on the
dummy integration variables.
Hence the replacements (\ref{10.1}) would have no effect on the
integrals themselves were it not for the fact that the integrands have
an explicit dependence on $x^0$ through the dependence of the action
functional (\ref{2.1}) on the volume element $d^4x$, which gets
multiplied by~${-i}$.
This dependence has the consequence that (\ref{5.1a}) and (\ref{5.4a})
get replaced by
\begin{eqnarray}
\left<\out\mid\mathop{\rm in}\right>&=&N\int
 e^{-S_E[\phi]}\mu[\phi]\,d\phi\,,\label{10.2}\\[\medskipamount]
\left<\out\right|A[\phi]\left|\mathop{\rm in}\right>&=&N\int A[\phi]
 e^{-S_E[\phi]}\mu[\phi]\,d\phi\,,\label{10.3}\\[\medskipamount]
\noalign{\noindent\rm where $S_E$ is the
{\sl Euclidean action},\medskip}
S_E[\phi]&=&{\textstyle{1\over 2}}\mu^2\int g_{ij}\left(\phi(x)\right)
\phi^i{}_{,\mu}\,\phi^j{}_{,\mu}\,d^4x\,,\label{10.4}
\end{eqnarray}
and the Greek indices now run from 1 to~4.
In eq.~(\ref{10.3}) there may also be additional dependence on $x^0$
and/or $p_0$ if $A[\phi]$, $\left|\mathop{\rm in}\right>$ or
$\left|\out\right>$ depends explicitly on one or both of these
variables.
In this case the corresponding dependence on $x^4$ and/or $p_4$ is
obtained by analytic continuation.

It will be noted in eq.~(\ref{10.4}) that the Minkowski metric
$\eta_{\mu\nu}$ has been replaced by the Euclidean~$\delta_{\mu\nu}$.
Moreover, in eq.~(\ref{10.3}) the chronological ordering symbol $T$ has
been dropped as irrelevant, the quantity on the left being now simply
regarded as {\sl defined\/} by the integral on the right.

In Minkowski spacetime the Fourier components of
$\phi^i(x)\left|\vac\right>$ behave like $e^{i\omega x^0}$ in the remote
past and those of $\left<\vac\right|\phi^i(x)$ behave like
$e^{-i\omega x^0}$ in the remote future (see eq.~(\ref{7.10})).
The corresponding behaviors in Euclidean space are $e^{\omega x^4}$ and
$e^{-\omega x^4}$ respectively, which imply that to get Euclidean vacuum
averages one must integrate, in eqs.~(\ref{10.2}) and (\ref{10.3}), over
fields that vanish at infinity.
On the computer, Euclidean space will be replaced by a \mbox{4-torus}
$T^4$, the fields will satisfy periodic boundary conditions, and only
their integrals over the \mbox{4-torus} will be required to vanish.
The corresponding ``Euclidean vacuum averages'' will be indicated by the
abbreviated notation.
\begin{equation}
\left<A[\phi]\right>={\int A[\phi]
e^{-S_E[\phi]}\mu[\phi]\,d\phi\over\int
e^{-S_E[\phi]}\mu[\phi]\,d\phi}\,.\label{10.5}
\end{equation}

It is not possible to use a Euclideanized version of the action
(\ref{2.5}) on a computer because a straightforward Euclideanization of
(\ref{2.5}) leads to an action that is not bounded from below.
One {\sl can\/} use a Euclideanized version for {\sl theoretical\/}
purposes if the Lagrange multiplier field $\lambda$ is itself subjected
to a rotation in the complex plane:
\begin{equation}
\lambda=-i\lambda_E\,.\label{10.6}
\end{equation}
A Euclideanized effective action
$\bar\Gamma_E\left[\bar\phi,\bar\lambda_E\right]$ can then be
introduced, as well as a corresponding $\Gamma_E\left[\bar\phi\right]$
having the form
\begin{eqnarray}
\Gamma_E\left[\bar\phi\right]&=&{\textstyle{1\over 2}}\bar\mu^2\int
\bar g_{ij}\left(\bar\phi(x)\right)\bar\phi^i{}_{,\mu}\,
\bar\phi^j{}_{,\mu}\,d^4x+\Delta
\Gamma_E\left[\bar\phi\right]\label{10.7}\\[\medskipamount]
\noalign{\noindent\rm(cf.~eq.~(\ref{8.15})).
It will be left as an exercise for the reader to rewrite section~8 in
Euclidean language.
We remark only that the renormalization constants all remain unchanged,
and equations~(\ref{8.17}) to (\ref{8.20}) take the forms\medskip}
\phi^a_R&=&Z^{-1}\phi^a\,,\label{10.8}\\[\medskipamount]
\Gamma_E&=&{\textstyle{1\over 2}}\mu^2_R\int g_{ij}
\left(\bar\phi_R(x)\right)\bar\phi^i_{R,\mu}\bar\phi^i_{R,\mu}\,d^4x+
\Delta\Gamma_E\,,\label{10.9}\\[\medskipamount]
\mu_R&=&Z\bar\mu\,,\label{10.10}\\[\medskipamount]
Z&=&\bar\phi^0=\left<\phi^0\right>\,.\label{10.11}
\end{eqnarray}

More generally, one may write
\begin{equation}
\mu_R=Z_\varphi\,\bar\mu_\varphi\label{10.12}
\end{equation}
where
\begin{equation}
Z_\varphi=\left<\left[f
\left(\,|\bmphi|\,\right)+\textstyle{1\over 2}\left|\bmphi\right|f'
\left(\,|\bmphi|\,\right)\right]\phi^0\right>\label{10.13}
\end{equation}
and where $\bar\mu_\varphi$ may be identified from the relation
\begin{equation}
\left<\varphi^i(x)\varphi^i(x')\right>={1\over(2\pi)^4}\int
{\delta_{ij}\,e^{ip\cdot(x-x')}\over\bar\mu_\varphi\,p^2+\cdots}
\,d^4p\,,\label{10.14}
\end{equation}
which follows from (\ref{7.7}) and the fact that, in the transition from
Minkowski spacetime to Euclidean space, the momentum volume element
$d^4p$ gets multiplied by~$i$.
In the special case in which the $\varphi^i$ are chosen to
be the Riemann normal coordinates $\sigma^i$
(see eqs.~(\ref{6.14}) and (\ref{6.15})) it is straightforward to verify
that
\begin{equation}
Z_\sigma={1\over 2}
\left<\left(1+{s\cosh s\over\sinh s}\right)\right>\,.\label{10.15}
\end{equation}
The variables $\sigma^i$ are particularly convenient to work with on the
computer, and in the following paper an account will be given of the
determination of $\mu_R$ based on the equation
\begin{eqnarray}
\left<\sigma^i_R(x)\sigma^j_R(x')\right>&=&{1\over(2\pi)^4}\int
{\delta_{ij}\,e^{ip\cdot(x-x')}\over\mu^2_Rp^2+\cdots}
\,d^4p\,,\label{10.16}\\[\medskipamount]
\sigma^i_R&=&Z^{-1}_\sigma\sigma^i\,.\label{10.17}
\end{eqnarray}

We note finally the Euclideanized versions of eqs.~(\ref{9.13}) and
(\ref{9.14}):
\begin{eqnarray}
&&\left<j_{0i\mu}(x)j_{0j\nu}(x')\right>=
{\mu^2_R\delta_{ij}\over(2\pi)^4}\int{p_\mu\,p_\nu\over p^2}
e^{ip\cdot(x-x')}\,d^4p
+\cdots\,,\qquad\mu\not=\nu\,,\label{10.18}\\[\medskipamount]
&&\left<\tilde\jmath_{0i\mu}(p)\tilde\jmath_{0j\nu}(p')\right>
=\mu^2_R\delta_{ij}{p_\mu\,p_\nu\over p^2}\delta(p+p')
+\cdots\,,\qquad\mu\not=\nu\,.\label{10.19}
\end{eqnarray}
The Fourier transforms are here defined as in eq.~(\ref{9.12}), and the
currents themselves are defined by
\begin{equation}
j_{ab\mu}(x)=\mu^2\left(\phi_a\,\phi_{b,\mu}
                       -\phi_b\,\phi_{a,\mu}\right)\label{10.21}
\end{equation}
(cf.~eq.~(\ref{4.17})).

\vspace{\bigskipamount}
This work has been supported in part by Grants PHY8919177, PHY9009850,
and\linebreak
PHY9120042 from the National Science Foundation, and by
the Robert A. Welch Foundation.
\clearpage
\renewcommand\theequation{A.\arabic{equation}} \setcounter{equation}{0}
\section*{Appendix}
Equations~(\ref{7.3}) and (\ref{7.5}) imply
\begin{equation}
\left<\vac\right|\varphi^i(x)\left|j,{\bf p}'\right>
 =i\bar\mu_\varphi\int_{-\infty}u(x',{\bf p}')
\left({\ora\ptl\over\ptl x'_\mu}-{\ola\ptl\over\ptl x'_\mu}\right)
\left<\vac\right|\varphi^i(x)\varphi^j(x')
\left|\vac\right>\,d\Sigma'_\mu\,.\label{a.1}
\end{equation}
This may be rewritten in the form
\begin{eqnarray}
&&{}-\bar\mu_\varphi
\left(\int_{+\infty}-\int_{-\infty}\right)u(x',{\bf p}')
\left({\ora\ptl\over\ptl x'_\mu}-{\ola\ptl\over\ptl x'_\mu}\right)
\left<\vac\right|T\left(\varphi^i(x)\varphi^j(x')\right)
\left|\vac\right>\,d\Sigma'_\mu\nonumber\\[\medskipamount]
&&{}=-i\bar\mu_\varphi\int u(x',p')
\left(\ora\square{}'-\ola\square{}'\right)
\left<\vac\right|T\left(\varphi^i(x)\varphi^j(x')\right)
\left|\vac\right>\,d^4x'\label{a.2}
\end{eqnarray}
since the insertion of the integral over the hypersurface ${+\infty}$
contributes nothing in view of (\ref{7.8}) and the kinematics of the
chronological product.
Inserting expressions (\ref{7.4}) and (\ref{7.7}) into (\ref{a.2}), and
using the fact that $u(x',p')\ola\square{}'=0$, one obtains
\begin{eqnarray}
\left<\vac\right|\varphi^i(x)\left|j,{\bf p}'\right>&=&(2\pi)^{-3/2}
\bar\mu_\varphi(2\omega')^{-1/2}\int d^4x'\int d^4p
e^{ip'\cdot x'}{1\over(2\pi)^4}\,{\delta_{ij}\,
e^{ip\cdot(x-x')}\over\bar\mu^2_\varphi+\cdots}\nonumber\\[\medskipamount]
&=&(2\pi)^{-3/2}\bar\mu_\varphi(2\omega')^{-1/2}\int{\delta_{ij}\,
\delta(p-p')e^{ip\cdot x}\over\bar\mu^2_\varphi+\cdots}
\,d^4p\nonumber\\[\medskipamount]
&=&\delta_{ij}\,\bar\mu^{-1}_\varphi u({\bf x},{\bf p}')\,.\label{a.3}
\end{eqnarray}
To reach the last line one makes use of the fact that the
\mbox{$\delta$-function} in the line above enforces the constraint
$p=p'$, $p^2=p'{}^2=0$, and hence the unwritten terms in the denominator
of the integrand vanish.

The adjoints of eqs.~(\ref{7.3}) and (\ref{7.5}) now yield
\begin{eqnarray}
\left<i,{\bf p}\mid j,{\bf p}'\right>&=&-i
\bar\mu_\varphi\int_{\pm\infty}u(x,{\bf p})^*
\left({\ora\ptl\over\ptl x_\mu}-{\ola\ptl\over\ptl x_\mu}\right)
\left<\vac\right|\varphi^i(x)
\left|j,{\bf p}'\right>\,d\Sigma_\mu\nonumber\\[\medskipamount]
&=&-i\delta_{ij}\int_{\pm\infty}
u(x,{\bf p})^*\left({\ora\ptl\over\ptl x_\mu}-{\ola\ptl\over\ptl x_\mu}\right)
u(x,{\bf p}')\,d\Sigma_\mu\nonumber\\[\medskipamount]
&=&\delta_{ij}\,\delta({\bf p}-{\bf p}')\,.\label{a.4}
\end{eqnarray}
\end{document}